\documentstyle[psfig,conf_iap]{article}
\begin{document}
\renewcommand{\textfraction}{0.1}
\renewcommand{\topfraction}{0.9}
\def\lesssim{\mathrel{\hbox{\rlap{\hbox{\lower5pt\hbox{$\sim$}}}\hbox{$<$}}}}
\heading{Metal systems of the Lyman forest}
%
\par\medskip\noindent
\author{%
Alec Boksenberg
}
\address{
University of Cambridge, Institute of Astronomy, Madingley Road,
Cambridge.  CB3 0HA, UK}

\begin{abstract}
An extremely careful analysis of the forest of weak metal-line systems
at $z \sim 2$ has allowed the separation into individual `single-phase
ionization' components with accurate parameters.  The systems
typically span several hundred \break km s$^{-1}$ and within each the
components show a strong coherence in their ionization properties.  A
general feature of the observed systems is the presence of high
ionization broad components which co-exist in velocity space with more
numerous, narrow, lower ionization components.  From a sample of
components extending to $z \sim 3.5$, taken either individually or as
system totals, no rapid evolution is found in the column density ratio
N(Si IV)/N(C IV), contrary to previous indication of a large change
near $z = 3.1$.  Comparisons of component ion ratios including Si IV,
${\rm C \ IV}$, C II and Si II from the sample at $z \sim 2$ with
CLOUDY derived values using several model ionizing spectra show the
general dominance of the metagalactic ionizing background over local
stellar radiation and give Si/C distributed 1--3 times the solar
value.  Indications of the evolution of structure are highlighted.
\end{abstract}
\section{Introduction }
With the spectra of exceptional quality delivered by the high
resolution spectrograph of the Keck telescope it has become possible
to detect individual metal features, most commonly the C IV $\lambda
\lambda$1548,1550 doublet, related to the high redshift Lyman forest
clouds for a large fraction of the stronger lines (Cowie {\it et
al}. 1996; Tytler {\it et al}. 1996; Womble, Sargent \& Lyons 1996;
Songaila \& Cowie 1996).  In general appearance such metal systems are
blended complexes of cloud components.  The deduced carbon abundance
in the complexes typically is $<10^{-2}$ of solar.  In their
full investigation Songaila \& Cowie deduce also that Si/C is about
three times solar (for each complex considered as an entity),
pointing to chemical abundances in these systems similar to
Galactic halo stars.  Additionally, they find rapid evolution in the
column density ratio N(Si IV)/N(C IV), from values of a few 
$10^{-1}$ at higher redshifts to about ten times lower at $z<3.1$
abruptly occurring in a redshift interval $<0.1$, which they suggest
arises from a sudden reduction in the opacity to He$^+$ ionizing
photons of the evolving intergalactic medium (see also Cowie in this
volume).

Here I give some new results on such metal systems from a full
analysis of a high quality spectrum of Q1626+643 $(z = 2.3)$
obtained on the Keck telescope at a resolution $\sim 6.3 $ km s$^{-1}$
FWHM, augmented with an analysis of similar quality Keck spectra
of Q1107+487 $(z = 3.0)$ and Q1422+231 $(z = 3.6)$.

\section{Kinematic Structure }

In Q1626+643 there are about 20 metal systems in the range $1.6 < z <
2.3$ as indicated by C IV absorption in the clear spectral region
between the Lyman$ \ \alpha$ and C IV emission lines.  These metal
complexes display a range of ionized species, where detectable in the
overall spectrum, from the weakest in which only C IV is detected to
the strongest which also contain Si IV, C II, Si II, ${\rm Al \ II}$, Al
III, O I, Fe II, Ni II and sometimes N V.  An example showing some
ions in a relatively rich system is in Figure 1.  Typically a complex
spans a few hundred km s$^{-1}$ and often is associated with one or
more others in a close group with wide expanses of clear spectrum
between such groups.

\begin{figure}
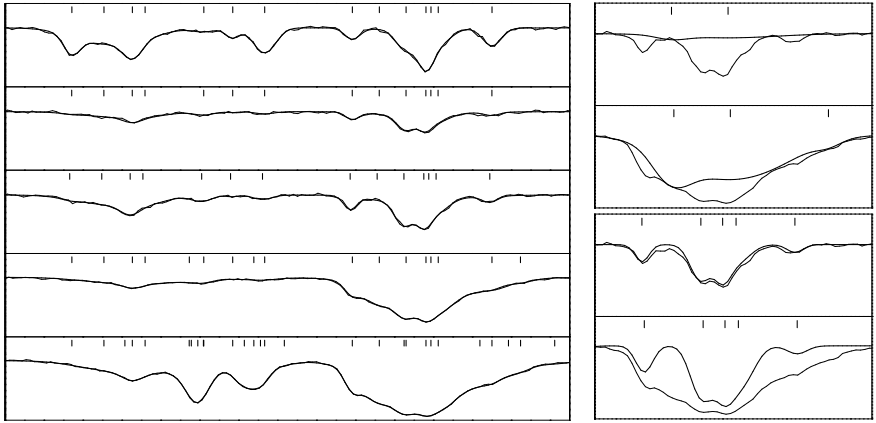

\vskip -1.9cm 
\hskip 0.4cm \hbox{\hbox{\psfig{figure=boksenbergF1a.ps,width=5.6cm}}
\hskip 2cm \vbox{\psfig{figure=boksenbergF1b.ps,width=2.76cm} \vskip
-0.925cm \psfig{figure=boksenbergF1c.ps,width=2.76cm}\vskip
-0.07cm}}
\caption{{\it Left} -- Complex at $z = 2.29$ in
the normalised spectrum of Q1626+643 and simultaneous profile fits
(bold smooth lines) superimposed, demonstrating the overall excellence
of the fits.  From the bottom: C IV $\lambda$1548 (overlying ${\rm C \ IV}$
$\lambda$1550 and weak Ni II $\lambda$1741 from other systems are
included), C IV $\lambda$1550, Si IV $\lambda$1393, Si IV
$\lambda$1402, C II $\lambda$1334.  The range shown spans 290 km
s$^{-1}$.  The ticks mark the positions of individual components, each
with appropriate $b$-value not indicated here. {\it Upper right} -- 
Deconstruction of a portion of the C IV $\lambda$1548 and Si IV
$\lambda$1393 profile fits, retaining only the high
ionization components shown in bold lines.  The range here spans 130 km
s$^{-1}$.  {\it Lower right} -- Similar, retaining only the lower ionization
components.}
\end{figure}

In an extremely careful analysis using the Voigt profile fitting
package VPFIT (e.g. Carswell {\it et al}. 1991) I have
self-consistently separated these complexes into individual component
clouds closely approximating single-phase ionization regions with
quite accurate values for column density, Doppler parameter $(b)$ and
redshift.  In the redshift range available to Si IV outside the Lyman
forest 68\% of the components defined in C IV are also detected in Si
IV.  Excellent simultaneous fits are achieved (as is obvious in Figure
1) with components each having the same redshift over all species, and
the same $b$-value for all ions of a given atom, but different column
densities.  Most of the components are narrow: 60\% have $b$(C IV)
distributed from 10 km s$^{-1}$ to below $4\ {\rm km \ s}^{-1}$ while 12\% have
$b > $ 20 km s$^{-1}$.  The distribution here is more peaked to lower
$b$-values than that found by Rauch {\it et al}. (1996) for clouds
dominantly at higher redshifts.  Particularly for the stronger
components the resultant ratios of $b$-values for ions of different
atoms, e.g., $b$(Si IV)/$b$(C IV), are physically realistic, and yield
a temperature structure within the complexes containing values
distributed up to a few $10^4$K, typical of photoionization heating,
as found earlier (Rauch {\it et al}. 1996).

A frequent feature of the systems is the presence of a few broad, high
ionization components which co-exist in velocity space with the much
more numerous narrow components of lower ionization.  A strong example
of this, indicated with C IV and Si IV, is shown in Figure 1.  In one
display the lower ionization components have been suppressed in the
fitted profile leaving a very broad component of $b$ = 32 km s$^{-1}$
with two narrower components, all of which are present strongly in C
IV but weak or absent in Si IV.  The converse case shows only the
lower ionization components, which now demonstrate a striking
similarity between the C IV and Si IV profiles.  Such sets of
different component structures must therefore be signatures of
physically distinct but spatially closely related regions which can be
isolated in the fitting procedure.  While the implicit model in such
profile constructions has only temperature and Gaussian turbulence
broadening included in the specific $b$-parameter characterising each
assumed cloud in a complex, large velocity gradients from bulk motions
must also contribute to the true overall absorption profile.
Nonetheless, any spectrum can be described with clustered Voigt
profile fits to the accuracy required with the available
signal-to-noise ratio.  The broad, high ionization components revealed
here thus probably represent regions of low volume density dominated
by bulk motions.  Comparison with the results of simulations
(e.g. Rauch, Haehnelt \& Steinmetz 1997) is important to derive proper
physical meaning from such observed profiles.

\section{Ionization Balance}

Figure 2 shows values of N(Si IV)/N(C IV) vs redshift for all system
components found in the spectrum of Q1626+643 having N(C IV) $\geq 6
\times 10^{11}\ {\rm cm}^{-2}$ and Si IV lying redward of the Lyman forest.
Immediately noticeable is a remarkable coherence in these values among
components in each system, typically extending over a factor only
$\sim 10$, while there are bulk differences between systems.  Such an
ionization pattern is not predicted in recent hydrodynamical
simulations of collapsing gas structures in the Universe photoionized
by a metagalactic ionizing background (Rauch, Haehnelt \& Steinmetz
1997), although in many other aspects this modelling shows
considerable success in reproducing observed characteristics of
absorbers.

\begin{figure}[t]
\vskip -1cm
\hbox{ \hskip -2.3cm 
\vbox{\psfig{figure=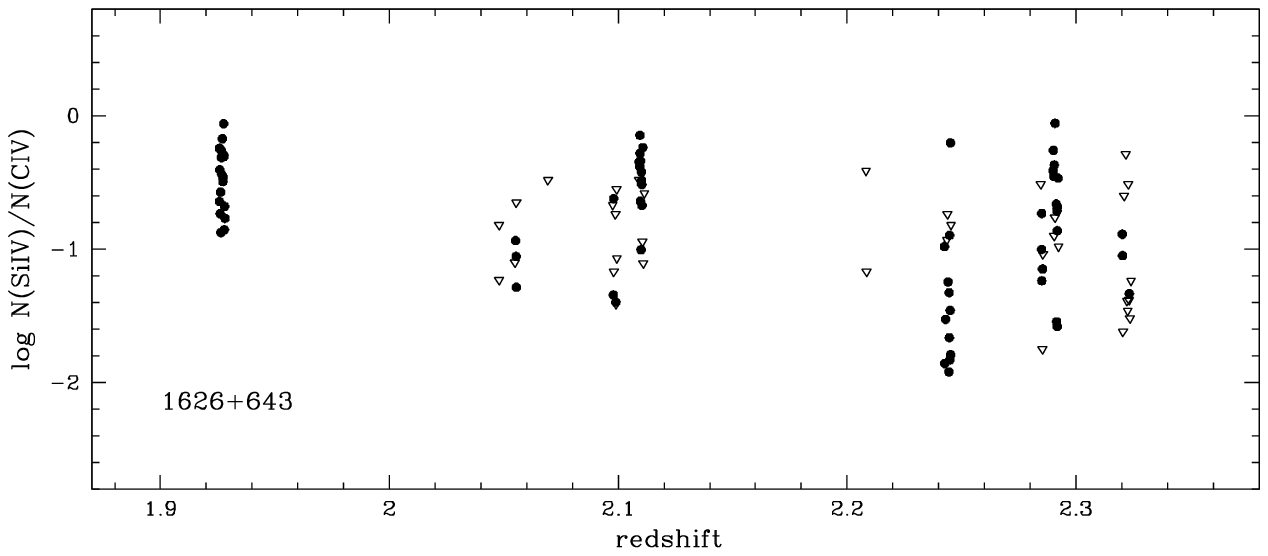,width=18cm}\vskip -12.5cm
\psfig{figure=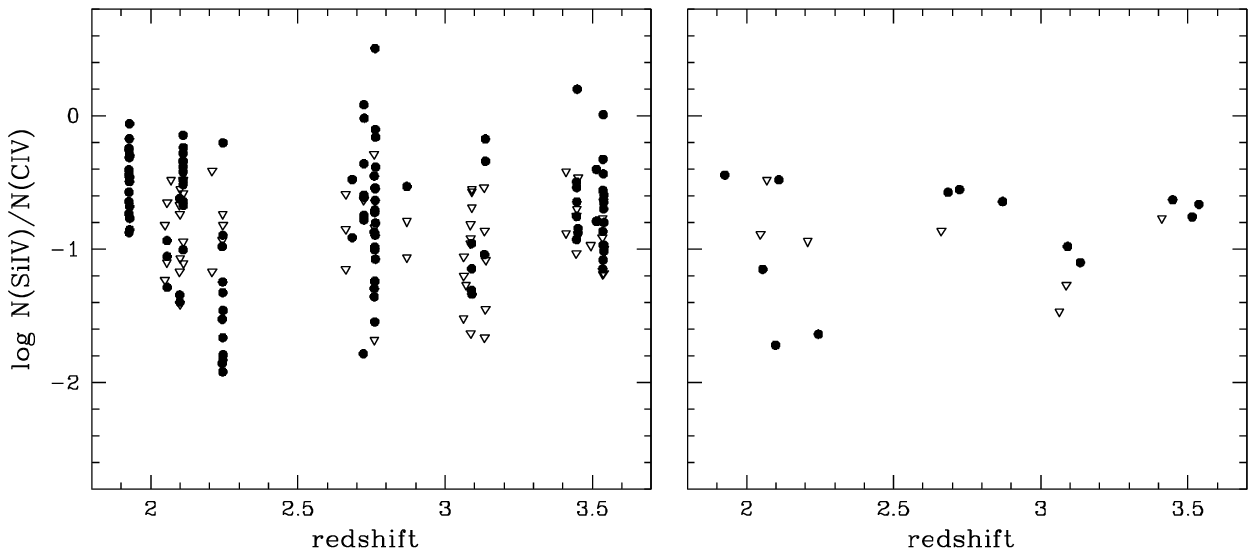,width=18cm}}}
\vskip -12cm
\caption{{\it Upper} -- N(Si IV)/N(C IV) vs redshift for all cloud
components in the spectrum of Q1626+643 with Si IV outside the Lyman
forest and ${\rm N(C \ IV)} \geq 6 \times 10^{11}$ cm$^{-2}$.  Inverted open
triangles are values with upper limits for ${\rm N(Si\ IV)}$.  To unify the
whole data set, where N(Si IV) $\leq 1.6 \times 10^{11}\ {\rm cm}^{-2}$
this value is taken as an additional upper limit. {\it Lower left} --
As upper panel but with an extended sample including components from
Q1107+487 and Q1422+231.  Here all components within 5000 km s$^{-1}$
of the QSO redshifts and systems assessed as not optically thin in the
Lyman continuum are excluded. {\it Lower right} -- As left but with
ratios of total system values.} 
\end{figure}

This data set is extended with equivalent high quality samples
analysed from the spectra of Q1107+487 and Q1422+231, giving coverage
over the range $1.9 < z < 3.5$; all are plotted in Figure 2, here
excluding systems within 5000 km s$^{-1}$ of the emission redshifts
and those not optically thin in the Lyman continuum as assessed from
the corresponding Lyman region spectra.  Contrary to the finding by
Songaila \& Cowie (1996) of rapid evolution near $z = 3.1$ there is no
significant overall change in the balance of the ratio values anywhere
over the whole range in redshift.  The use here of single-phase
ionization cloud components to probe the radiation field must be more
effective than summing over whole systems as done by Songaila \&
Cowie, but to compare with their data the ratios of the separately
totalled system values also are plotted in Figure 2: again there is no
evidence for strong evolution near $z=3.1$, although at $z \sim 2$ the
scatter in these values seems larger than at higher redshifts.  This
may be indicating the evolution of collapsing structures.  An evolving
$b$-value distribution, as noted above, and evident change, e.g., in
N(C II)/N(C IV) indicating a trend to higher densities at lower
redshifts, also point to this.  Such structure evolution will be
investigated fully in a later paper (in preparation).

\begin{figure}[t]
\hbox{\psfig{figure=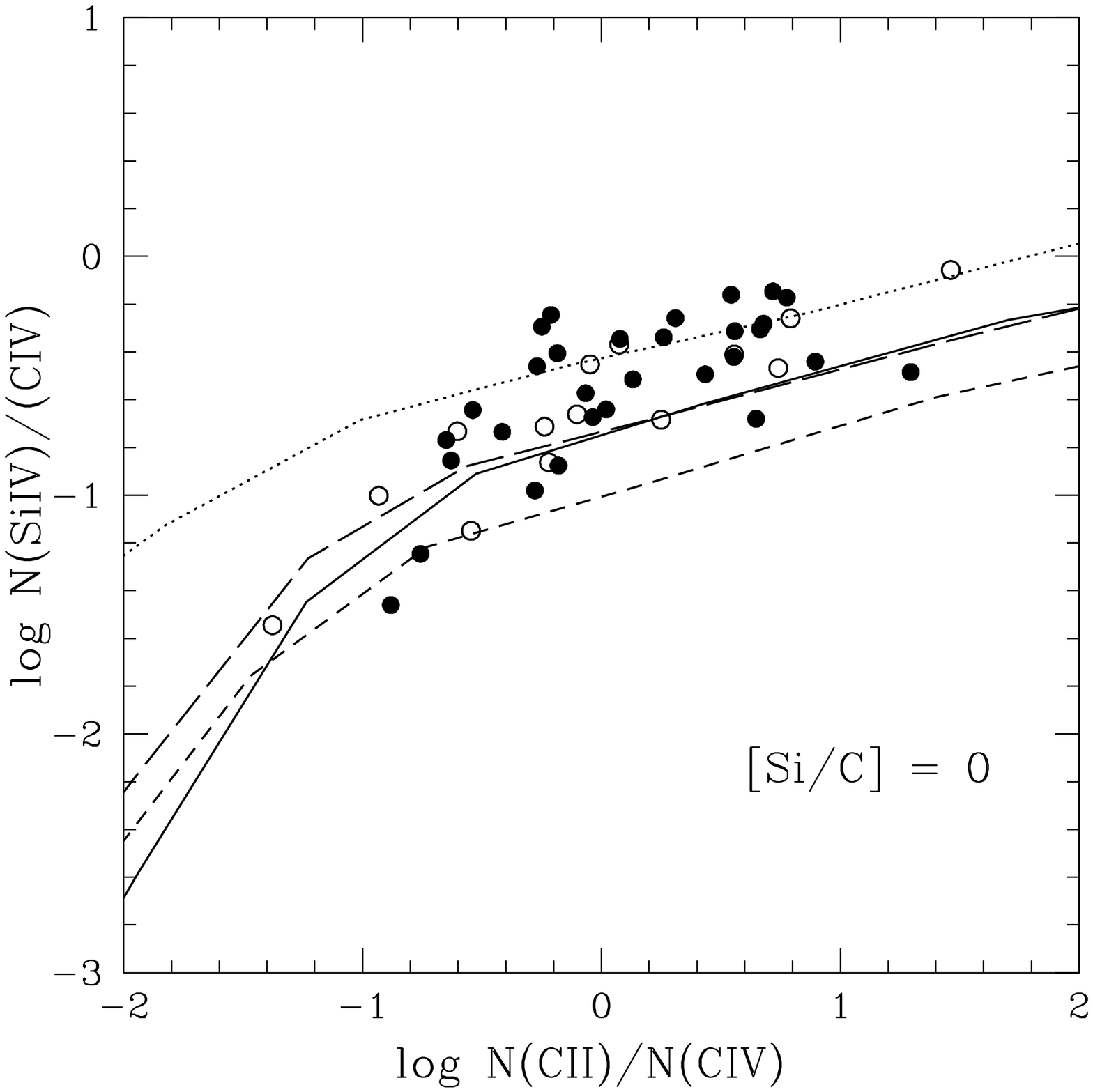,width=5.9cm} \hskip 0.1cm
\psfig{figure=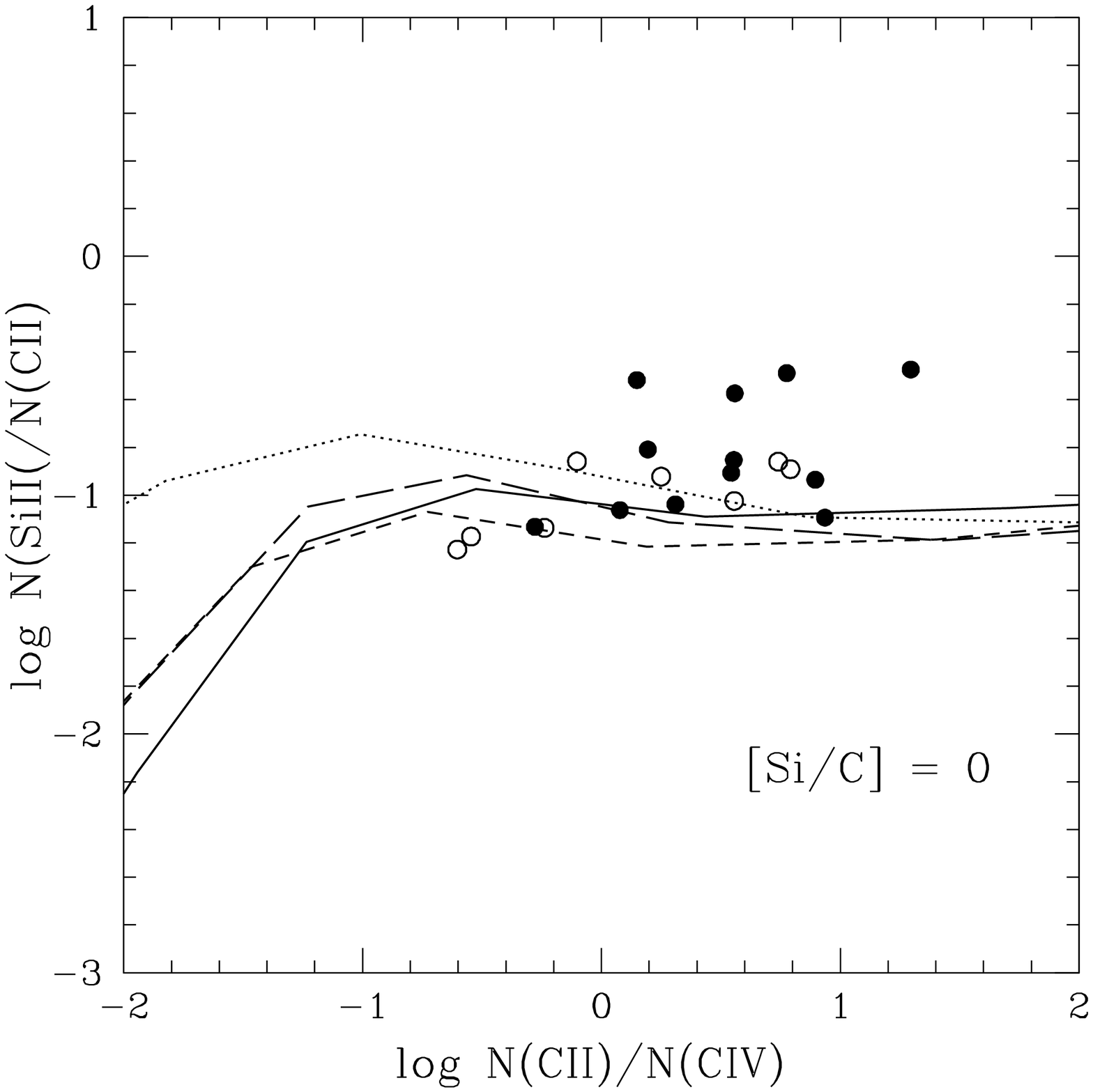,width=5.9cm}}
\caption{Comparison of component ion column density ratios at $z \sim
2$ from the spectrum of Q1623+643 with model predictions of the CLOUDY
code in the optically thin regime and for low metallicity.  Open symbols show data within 5000 km s$^{-1}$ of the QSO.  The
models are computed for: (a) Haardt \& Madau (1996) latest available
versions for the UV background (Haardt 1997) using $q_0 = 0.5$ and $z
= 2.160$, and showing two source cases: QSOs -- solid line; QSOs +
galaxies -- long-dash line; (b) power law of $\nu^{-1.8}$ --
short-dash line; a 5000K, log $g = 4.5$ stellar spectrum (Kurucz 1979)
available within the CLOUDY code -- dotted line.  The cosmic microwave
background, a significant cause of Compton cooling for low density
clouds, is included in all cases.}
\end{figure}

More information on the shape of the ionizing spectrum can be obtained
from displays of the ratios N(Si IV)/N(C IV) vs N(C II)/N(C IV) than
from N(Si IV)/N(C IV) alone (Songaila \& Cowie 1996; Giroux \& Shull
1997).  In Figure 3 is such a plot from the Q1626+643 data set
compared with model predictions of the CLOUDY code (Ferland 1996) for
several assumed ionizing backgrounds and for solar relative abundance
of Si/C.  For the broadly distributed material (filled symbols),
evident is the relative insensitivity to the spectral shape assumed
for the metagalactic background in the span of most of the data.  This
allows component values for Si/C distributed up to about three times
solar, the level previously indicated at higher redshifts for system
totals (Songaila \& Cowie 1996; Rauch, Haehnelt \& Steinmetz 1997)
although for clouds with log N(C II)/N(C IV) $\lesssim -0.3$ (i.e., at
lower densities) values seem closer to solar.  The same conclusion
comes for clouds close to the QSO (open symbols), here using the more
appropriate pure power law spectrum.  On the other hand, if the
ionizing spectrum is dominated by local effects from massive stars,
higher values of N(Si IV)/N(C IV) will result (Giroux \& Shull 1997;
Savaglio {\it et al}. 1997) and Si/C may be closer to the solar value
throughout, as illustrated here with use of a 50000K star model
spectrum.  An alternative display using the same model ionizing
spectra to construct the ratios ${\rm N(Si\ II)}$/N(C II) vs N(C II)/N(C IV),
also shown in Figure 3, gives less overall dependency on spectral
shape so is a better indicator of relative abundance, again in the
span of most of the data.  A similar distribution for Si/C again results.
Taken together, the two representations indicate that local stellar
sources do not dominate the radiation environment for these absorbers,
although the not complete consistency between them suggests some, not
unexpected, departure from strict photoionization equilibrium (e.g.,
Rauch, Haehnelt \& Steinmetz).  This is developed in a more complete investigation including
several other ions in a later paper (in preparation).

\acknowledgements{I thank Wal Sargent for his generous offer to
collaborate with him on the work on the Q1626+643 spectrum and thank
him and Michael Rauch for extension of this to include the Q1107+487
and Q1422+231 spectra, all of which are the subject of fuller papers in
preparation.  I also thank Martin Haehnelt for very helpful
discussions and illuminating new material concerning results of
hydrodynamical simulations.  I am greatly indebted to Bob Carswell for
providing VPFIT and spending so much time schooling me in its use. }



\vfill
\end{document}